\documentclass[aps,pra,amsmath,amssymb,twocolumn,showpacs,superscriptaddress]{revtex4}
\usepackage{bbm}
\usepackage{graphicx}
\usepackage{amsmath}
\usepackage{amssymb,amsthm}
\usepackage{color}
\usepackage{hyperref}
\usepackage[normalem]{ulem}

\newcommand{\eq}{\begin{eqnarray}}
\newcommand{\en}{\end{eqnarray}}
\def\bra#1{\mathinner{\langle{#1}|}}
\def\ket#1{\mathinner{|{#1}\rangle}}
\newcommand{\braket}[2]{\langle #1|#2\rangle}

\begin{document}

\title{Statistical signatures of states orthogonal to the Fock-state ladder of composite bosons}

\author{P. Alexander Bouvrie}
\affiliation{Centro Brasileiro de Pesquisas F\'isicas, Rua Dr. Xavier Sigaud 150, Rio de Janeiro, RJ 22290-180, Brazil}

\author{Malte C. Tichy}
\affiliation{Department of Physics and Astronomy, University of Aarhus, DK-8000 Aarhus C, Denmark}

\author{Klaus M\o{}lmer}
\affiliation{Department of Physics and Astronomy, University of Aarhus, DK-8000 Aarhus C, Denmark}

\pacs{
05.30.-d, 
05.30.Fk 
05.30.Jp, 
03.65.Ud 
}

\date{\today}

\begin{abstract}
The theory of composite bosons (cobosons) made of two fermions [Phys. Rev. A {\bf 71}, 034306 (2005), Phys. Rev. Lett. {\bf 109}, 260403 (2012)] converges to ordinary structureless bosons in the limit of infinitely strong entanglement between the fermionic constituents. For finite entanglement, the annihilation operator $\hat c$ of a composite boson couples the $N$-coboson Fock-state not only to the $(N-1)$-coboson state -- as for ordinary bosons --, but also to a  component which is orthogonal to the Fock-state ladder of cobosons. Coupling with states orthogonal to the Fock ladder arises also in dynamical processes of cobosons. Here, with a Gedanken-experiment involving both mode-splitting and collective Hong-Ou-Mandel-like interference, we derive the characteristic physical signature of the states orthogonal to the Fock ladder generated in the splitting process. This allows us to extract microscopic properties of many-fermion-wave functions from the collective coboson behavior. We show that consecutive beam-splitter dynamics increases the deviation from the ideal bosonic behavior pattern, which opens up a rigorous approach to the falsification of coboson theory.
\end{abstract}

\maketitle

\section{Introduction}
The exchange anti-symmetry of fermionic wave functions implies the Pauli exclusion principle, which in the most fundamental way governs the properties of atoms, molecules and solids, while the exchange symmetry of bosonic wave functions leads to lasing, Bose-Einstein condensation and statistical correlations that can be observed in interference processes with photons \cite{HongOuMandel1987} and with bosonic atoms \cite{LopesImanalievEtal2015}. Consistent with the spin-statistics theorem \cite{fierz1939,Pauli1940}, systems comprised of an even number of fermions can be treated as composite bosons \cite{Atkins1974}. However, the commutation relations for composite bosons are modified by their underlying structure \cite{CombescotLeyronasEtal2003,Law2005}. Strong binding of the constituent fermions is required to ensure ideal bosonic behavior, but, at the formal level, it is not the binding but the {\it entanglement} between the constituents that warrants the bosonic properties of composite systems \cite{Law2005,ChudzickiOkeEtal2010}. This has been exemplified by studying how varying entanglement lead to larger or smaller deviations from ideal bosonic behavior both in static many-body properties \cite{RombutsVanNeck2002,Law2005,PongLaw2007,ChudzickiOkeEtal2010,CombescotLeyronasEtal2003,CombescotMatibetEtal2008,CombescotBubinEtal2009,CombescotMatibet2010,CombescotShiauEtal2011,Combescot2011,RamanathanKurzynski2011,LeeThompsonEtAl2013,LeeThompsonEtall2014,GavrilikMishchenko2012,Gavrilik2013,TichyBouvrie2012a,TichyBouvrie2013,TichyBouvrie2014,PRLCombescot2015} and in dynamical processes \cite{BroughamBarnett2010,TichyBouvrie2012b,KurzynskiRamanathan1012,Thilagam2013,Thilagam2015,CombescotShiauChang2016}.

In this article, we consider composite bosons, {\it cobosons}, made up of two distinguishable fermions, $a$ and $b$ \cite{RombutsVanNeck2002,Law2005,PongLaw2007,ChudzickiOkeEtal2010,CombescotLeyronasEtal2003,CombescotMatibetEtal2008,CombescotBubinEtal2009,CombescotMatibet2010,CombescotShiauEtal2011,Combescot2011,RamanathanKurzynski2011,LeeThompsonEtAl2013,GavrilikMishchenko2012,Gavrilik2013,TichyBouvrie2012a,TichyBouvrie2014,LeeThompsonEtall2014,PRLCombescot2015,BroughamBarnett2010,TichyBouvrie2012b,KurzynskiRamanathan1012,Thilagam2013,Thilagam2015,CombescotShiauChang2016}. The wave function of one coboson is represented in second quantization by the action of the corresponding coboson creation operator $\hat c^\dagger$ on the vacuum, $\ket{1} = \hat c^\dagger \ket{0}$. The successive application of the creation operator defines the Fock-states ladder of cobosons
\eq
\label{NCobState}
\ket{N} = \frac{\left(\hat c^\dagger \right)^N}{\sqrt{ N! \chi_{N}} } \ket{0},
\en
where $\chi_{N}$ is the $N$-coboson {\it normalization factor} \cite{CombescotLeyronasEtal2003}. The coboson theory possesses peculiarities which differ at the fundamental level from those of elementary particles. For instance, the $\ket{N}$ are not eigenstates of the operator $\hat c^\dagger \hat c$, in contrast to ideal bosonic Fock-states. This is rooted in the fact that the application of the annihilation operator to any of these states with $N > 1$ populates a  component outside the ladder \cite{Law2005,ChudzickiOkeEtal2010,KurzynskiRamanathan1012},
\eq
\label{cvecN}
\hat c \ket{N} = \sqrt{\frac{\chi_{N}}{\chi_{N-1}}}\sqrt{N}\ket{N-1}+\ket{\varepsilon_{N-1}},
\en
where $\ket{\varepsilon_{N-1}}$ describes $N-1$ fermion pairs but is orthogonal to $\ket{N-1}$ and to any $\ket{\varepsilon_i}$ with $i \neq N-1$. In the literature, no attention has been paid to the physical meaning of $\ket{\varepsilon_{N-1}}$ and their possible observable consequences. The state $\ket{\varepsilon_{N-1}}$ entails an intricate, but largely disregarded, commutator algebra \cite{CombescotBubinEtal2009} e.g., in the construction of the coboson Kraus operator \cite{KurzynskiRamanathan1012} and coherent states of cobosons \cite{LeeThompsonEtAl2013}.

Nevertheless, dynamical processes of cobosons with finite entanglement between their constituent fermions  inevitably cause transitions into states orthogonal to the Fock ladder \cite{Law2005}, which may lead to observable consequences in the particle statistics. Here, we show that linear mode splitting of cobosons generates states of the form $\ket{M,N-M}^\perp$, which are orthogonal to the two-mode Fock ladder $\ket{M,N-M}$, and that their impact on Hong-Ou-Mandel  (HOM) like counting statistics \cite{HongOuMandel1987} in a post selection process involving collective interference is not negligible. Non ideal bosonic behavior is reflected by the nonideal collective HOM-like interference of two  Fock-states of cobosons \cite{TichyBouvrie2012b}, but when the states $\ket{M,N-M}^\perp$ come into play in the interference, the deviation from the ideal bosonic pattern can increase even more. In particular, we find changes by orders of magnitude for a wide range of strong entanglement, which facilitates the search for non-ideal bosonic signatures of composite bosons.

The article is organized as follows: In Sec.~\ref{CobFormalism} we summarize elements of the formalism for many-coboson states. In Sec.~\ref{CobSplitting} we consider the splitting of the system as implemented in a 2D lattice model where $\ket{a_j,b_j}$ denotes a fermion pair occupying the $j$th lattice site, which is allowed to cotunnel into the $j$th site in a parallel lattice. Such a scenario is feasible in experiments with attractively interacting fermionic atoms in tunable potentials \cite{SerwaneZurnEtal2011,ZurnSerwane2012,ZurnJochim2013,Jochim2015}. In Sec.~\ref{DoubleBS}, we introduce further mode-mixing with a third lattice, which is initially occupied by a single coboson, and we show that such a setup, by post selection, offers an amplification of the signatures of the composite character of the cobosons. Section ~\ref{Conclusion} concludes the work.

\section{Composite boson formalism}
\label{CobFormalism}

In this section, we summarize the main ingredients of the theory of composite bosons made up of two fermions \cite{CombescotMatibetEtal2008,CombescotLeyronasEtal2003,Law2005,ChudzickiOkeEtal2010,TichyBouvrie2012a}. Consider a composite boson made of two distinguishable fermions, $a$ and $b$, with a wave function of the form
\eq
\label{CobosonWF}
\ket{\psi}= \sum_{i,j=1}^\infty w_{i,j} \ket{A_i,B_j},
\en
where $\{\ket{A_i}\}$ and $\{\ket{B_j}\}$ are complete bases of single-fermion states. Due to the {\it Schmidt decomposition} \cite{Bernstein2009}, we can identify new bases, $\{\ket{a_j}\}$ and $\{\ket{b_j}\}$, such that
\eq
\label{CobosonWFSchmidt}
\ket{\psi}= \sum_{j=1}^S \sqrt{\lambda_j} \ket{a_j,b_j}.
\en
The distribution $\Lambda=\{\lambda_1,\ldots,\lambda_S\}$ of the Schmidt coefficients $\lambda_j$ associated with the two-fermion states $\ket{a_j,b_j}$ characterizes the entanglement of $\ket{\psi}$, via the purity of either reduced single-fermion density-matrix,
\begin{equation}
\label{eq:purity}
P=\sum_{j=1}^S \lambda_j^2.
\end{equation}

Fermion pairs in the composite-boson state $\ket{\psi}$ (Eqs.(\ref{CobosonWF}) and (\ref{CobosonWFSchmidt})) naturally motivate the coboson creation operator  \cite{Law2005}
\eq
\label{CreationOp}
\hat c^\dagger=\sum_{j=1}^S \sqrt{\lambda_j}  \hat a^\dagger_j \hat b^\dagger_j =\sum_{j=1}^S \sqrt{\lambda_j}  \hat d^\dagger_j ,
\en
where $\hat d^\dagger_j \equiv \hat a^\dagger_j$ $\hat b^\dagger_j$ creates a pair of fermions, a {\it bifermion}, in the product state $\ket{a_j,b_j}$ in the $j$-th Schmidt mode. The composite boson normalization factor $\chi_{N}$ (see Eq.~\eqref{NCobState}) reflects how, to obey the Pauli principle, the fermion pairs must distribute themselves avoiding multiple occupation of any of the $S$, possibly $\infty$, Schmidt modes. It can be evaluated as \cite{Law2005}
\eq
\chi_{N}^{\Lambda} = N! \sum^S_{p_1 < p_2 < \cdots < p_N} \lambda_{p_1} \lambda_{p_2} \cdots \lambda_{p_N},
\en
which constitutes a particular case of the elementary symmetric polynomial \cite{CombinatoryAnalysisMacmahon,TichyBouvrie2012a,TichyBouvrie2014}. We omit the index $\Lambda$ in $\chi_N^{\Lambda}$ unless necessary to specify a distribution of coefficients different from $\Lambda$.

Past works have studied how the operator $\hat{c}^\dagger$ deviates from the canonical bosonic creation operator when applied multiple times to the vacuum. This deviation can be traced back to the Pauli blocking of the fermionic constituents and the resulting modification of the commutation relations of $\hat{c}^\dagger$ and its adjoint annihilation operator, $\hat{c} = \left(\hat{c}^\dagger\right)^\dagger$. The expectation value of the commutator \cite{Law2005}
\eq
\label{ccdagger}
[\hat c ,\hat c^\dagger] = 1-\Delta ,
\en
with
\eq
\Delta = \sum_{j=1}^S \lambda_j (\hat a_{j}^\dagger \hat a_{j} + \hat b_{j}^\dagger \hat b_{j}),
\en
in the state $|N\rangle$ is given by \cite{ChudzickiOkeEtal2010}
\eq
\label{commutatorexpl}
\bra N  \left[ \hat c, \hat c^\dagger \right] \ket N = 2 \frac{\chi_{N+1}}{\chi_{N}} -1 .
\en
Perfect bosonic behavior is obtained when the normalization factors equal unity for all $N$, which occurs in the limit in which the fermion pairs do not compete for single-fermion states, i.e., when the coboson is a highly entangled ($P\approx0$) pair of fermions that distributes over many Schmidt modes with small occupation probability, $\lambda_j\approx0$. The {\it normalization ratio} $\frac{\chi_{N+1}}{\chi_{N}}$ captures quantitatively the bosonic quality of $N$ cobosons prepared in  $\ket{N}$. For a given purity $P$, there are two particular Schmidt distributions which maximize and minimize the normalization ratio for a given purity $P$ \cite{TichyBouvrie2012a,TichyBouvrie2014}, namely, the peaked $\Lambda^\text{peak}$ and the uniform $\Lambda^\text{uni}$ distributions, respectively. The upper and the lower bounds on $\chi_N$ in $P$ converge to $1$ in the limit $P\rightarrow 0$. Deviations of the normalization ratio from unity entail observable consequences \cite{CombescotMatibetEtal2008,CombescotBubinEtal2009,CombescotShiauEtal2011,KurzynskiRamanathan1012,TichyBouvrie2012b,LeeThompsonEtAl2013,LeeThompsonEtall2014,PRLCombescot2015} and occur equally well for cobosons made of bosonic constituents \cite{TichyBouvrie2013}.

While the successive application of $\hat c^\dagger$ to the vacuum leads to a discrete ladder of many-body states, the application of the annihilation operator to $\ket{N}$ populates a state component, $\ket{\varepsilon_{N-1}}$, which is orthogonal to the Fock-state ladder, \eqref{cvecN}. The  component $\ket{\varepsilon_{N-1}}$ has norm \cite{Law2005},
\eq
\braket{\varepsilon_{N-1}}{\varepsilon_{N-1}} = 1 - N \frac{\chi_{N}}{\chi_{N-1}}  + (N-1) \frac{\chi_{N+1}}{\chi_{N}},
\en
which vanishes for unit normalization factors (perfect bosonic behavior), while for finite purity (\ref{eq:purity}) the normalization factors differ from unity and the contribution of $\ket{\varepsilon_{N-1}}$ comes into play.

Due to the exact mapping of hardcore-bosons to bifermions, the dynamical properties of strongly bound fermion pairs $\hat d_{j}^\dagger$ (bifermions) in a lattice can be simulated by hardcore-bosons $\hat h_{j}^\dagger$, and vice versa. Indeed, the equivalent many-body ladder of Fock-states \eqref{NCobState} for hardcore-bosons, given by  $\ket{N}_{\rm h.c.} = (\hat B^\dagger )^N/\sqrt{ N! \chi_{N}} \ket{0}$ with $\hat B^\dagger = \sum_{i=1}^S \sqrt{\lambda_j} \hat h_j^\dagger$, shares the same features as the coboson-subtracted state, \eqref{cvecN}, when the corresponding anihilator operator $\hat B = \sum_{i=1}^S \sqrt{\lambda_j} \hat h_j$ acts on it, as well as a similar commutation relation, $[\hat B ,\hat B^\dagger] = 1-\Delta_{\rm h.c.}$, where $\Delta_{\rm h.c.}= 2 \sum_{j=1}^S \lambda_j \hat h_j^\dagger \hat h_j$.

\section{Mode splitting of composite bosons}
\label{CobSplitting}

\begin{figure}[ht]
\vspace{0.5cm}
\includegraphics[scale=0.23]{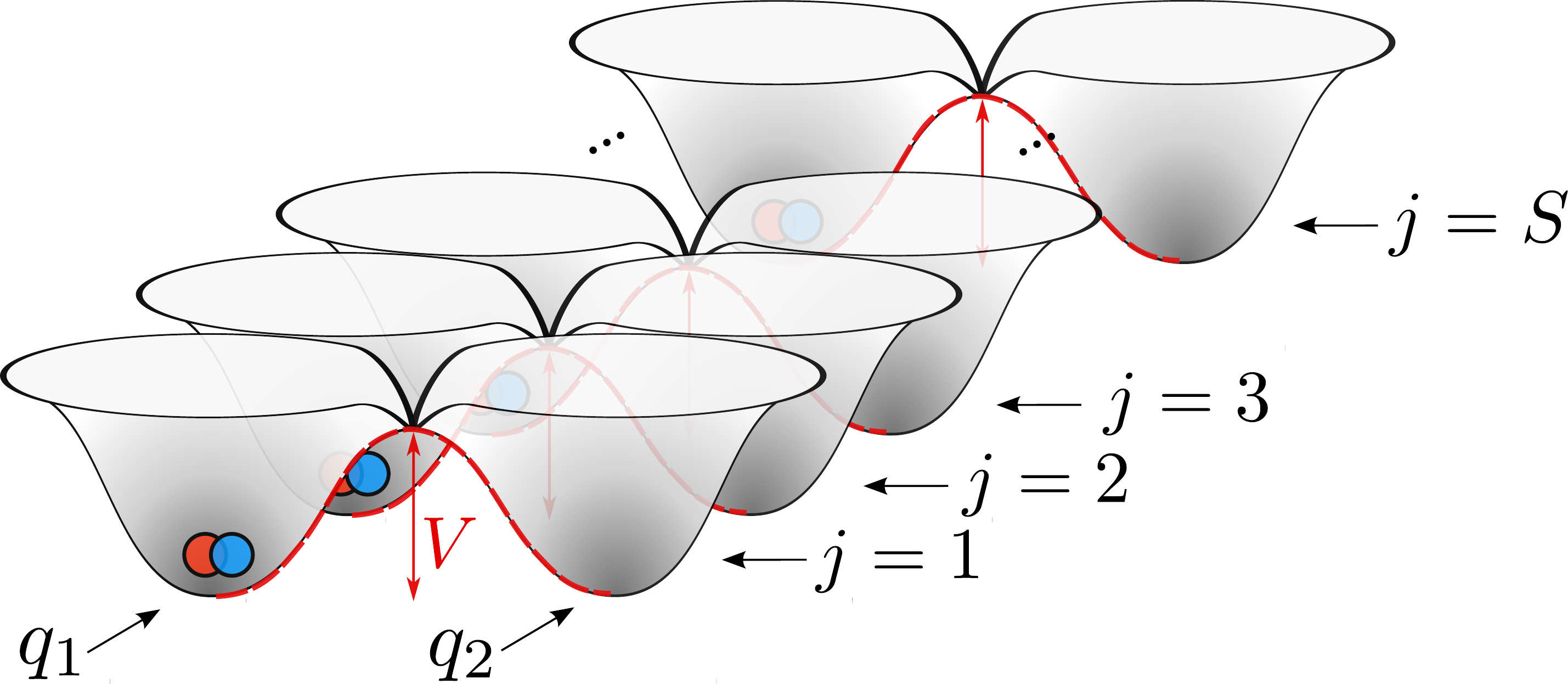}
\caption{Two-dimensional lattice setup for coboson splitting processes. $N$ strongly bound bifermions are prepared in the first sublattice ($q=1$). The barrier between the first and the second sublattices ($V\propto 1/J$ ) is then ramped down, allowing tunneling between the $j$th wells of the sublattices. A $50/50$ beam-spitter-like dynamics is obtained after a time evolution $t = \pi /(2J)$, where $J$ is the tunneling amplitude.}
\label{SplittingSchedule}
\end{figure}

In Ref.~\cite{TichyBouvrie2012b}, a system of strongly bound fermion pairs that are trapped in a two-dimensional potential landscape \cite{StrzysAnglin2010} was proposed to imitate the interference process of two-fermion composite bosons. The model, see Fig.~\ref{SplittingSchedule}, consists of two parallel sublattices in which pairs of ultracold fermionic atoms in the strong binding regime \cite{Jochim2015} are allowed to cotunnel between the $j$th wells of the sublattice but not between wells along the sublattice direction. The sublattices are identified as the two (external) input/output modes of the beam-splitter, and the $S$ wells of each sublattice as the (internal) Schmidt modes. The experimental protocol to prepare the initial two-mode coboson state, $\ket{N_1,N_2}$, is provided in the supplemental material of \cite{TichyBouvrie2012b}. However, although the resulting HOM-like counting statistics in the coboson interference was computed exactly, no attention was devoted to the statistics of the constituents, and an exact description of the resulting collective wave function was not provided. In this section we address these issues for the simplest case, namely, the splitting of an $N$ coboson Fock-state.

\subsection{Mode splitting dynamics}

We start with $N$ cobosons prepared in the first horizontally extended lattice, $q=1$, and an empty second sublattice, $q=2$, that is,
\eq
\label{InState}
\ket{N,0} = \frac{\left(\hat c_1^\dagger\right)^N}{\sqrt{N! \chi_{N}}} \ket{0,0},
\en
where $\hat c_q^\dagger = \sum_{j=1}^{S} \sqrt{\lambda_j} \hat d^\dagger_{q,j}$. Since, in the splitting process, there is at most a single bifermion in the $j$th pair of wells, on-site bifermion interactions can be neglected, and thus, the Pauli principle allows us to describe the subsequent dynamics by the operator evolution
\begin{equation}
\label{OperatorDynamics12}
\hat d_{1,j}^\dagger \longrightarrow \left( \sqrt{R} \hat d_{1,j}^\dagger + \sqrt{T} \hat d_{2,j}^\dagger \right),
\end{equation}
where $T=\sin^2(J t/2)$ and $R=\cos^2(J t/2)$ are the transmission and reflection probabilities and $J$ is the coupling (tunneling) amplitude between the $j$th modes of the two sublattices. The time evolution until $t=\pi/(2J)$ implements a balanced beam-splitter with $T=R=1/2$.

\subsection{State after a cobosons splitting}
\label{SplittingAndState}

Using the evolution operator \eqref{OperatorDynamics12} for a balanced beam-splitter, the initial state $\ket{N,0}$ evolves as

\begin{widetext}
\eq
\label{FiStateA}
\ket{N,0} \longrightarrow \ket{\Psi_{N}} &=& \frac{1}{\sqrt{2^N N! \chi_{N}}} \sum_{\substack{k_1,\ldots,k_{N}=1 \\ k_1 \neq \cdots \neq k_{N}}}^S \prod_{i=1}^N \sqrt{\lambda_{k_i}} ~ \left( \hat d_{1,k_i}^\dagger + \hat d_{2,k_i}^\dagger \right) \ket{0,0}. \nonumber \\
&=& \sqrt{\frac{1}{2^N N! \chi_{N}}} \sum_{M=0}^N\binom{N}{M} \sum_{\substack{k_1,\ldots,k_{N}=1 \\ k_1 \neq \cdots \neq k_{N}}}^S \prod_{i=1}^M \sqrt{\lambda_{k_i}} \hat d_{1,k_i}^\dagger \prod_{j=M+1}^{N} \sqrt{\lambda_{k_j}} \hat d_{2,k_j}^\dagger \ket{0,0}.
\en
\end{widetext}

\begin{figure}[ht]
\centering
\includegraphics[scale=0.7]{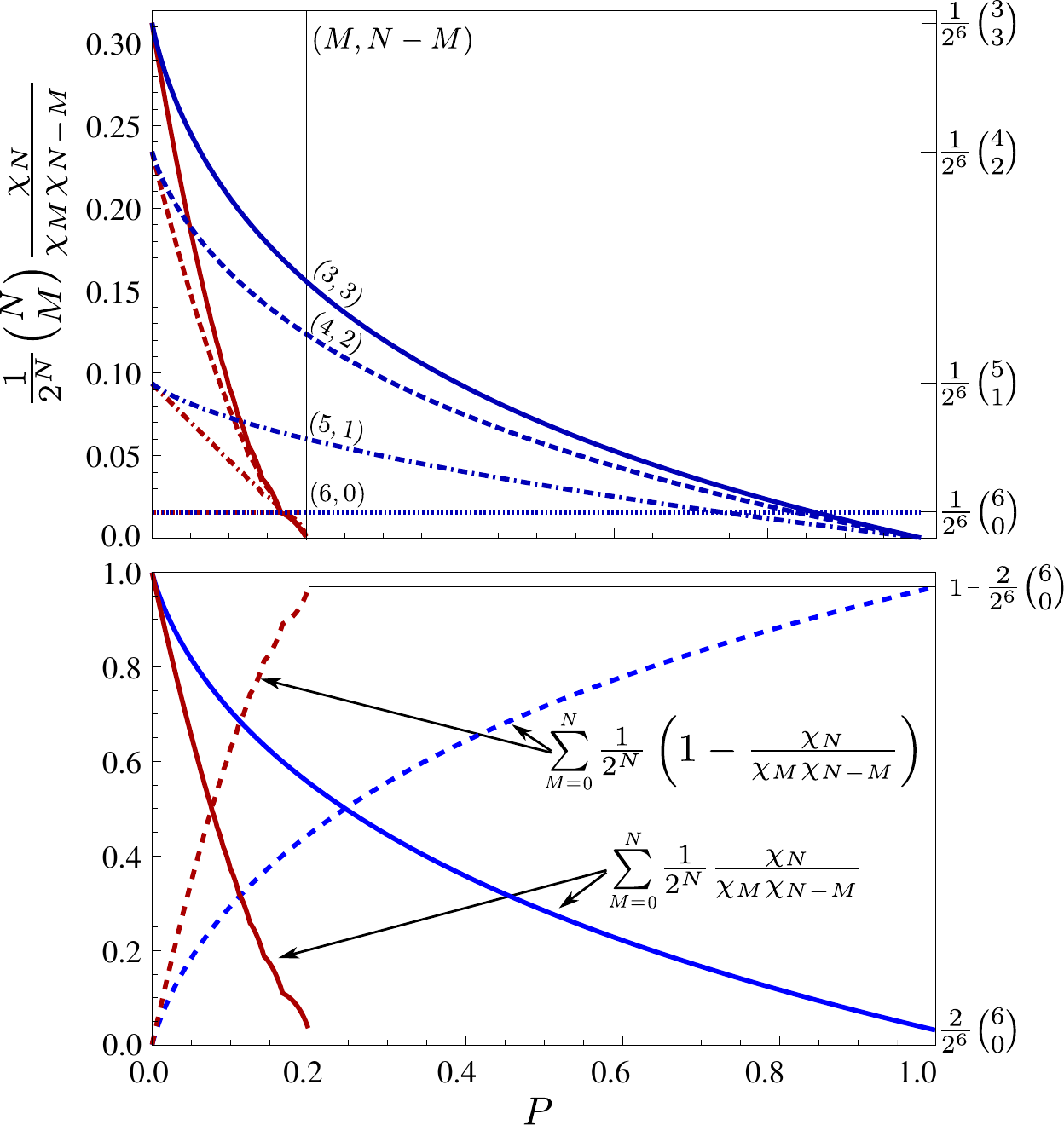}
\caption{(Color online) Population of the coboson component, cf. \eqref{FinStateWithOrthogonals} for $N=6$ as a function of the purity (upper panel). Blue lines correspond to the peaked distribution $\Lambda^\text{peak}$ of Schmidt coefficients and red lines (up to $P=1/5$) to the uniform distribution $\Lambda^\text{uni}$ \cite{TichyBouvrie2012a}. These distributions maximize and minimize the normalization ratio $\frac{\chi_N}{\chi_{N-1}}$ and, hence, $\frac{\chi_N}{\chi_M \chi_{N-M}}$ for a given $P$. The lower panel displays the sum over all contributions of finding $M$ and $N-M$ bifermions in the outputs in a proper coboson state (solid lines) or in the states $\ket{M,N-M}^\bot$ (dashed lines), cf. \eqref{FinStateWithOrthogonals}.}
\label{CobAndPerpStatistics}
\end{figure}

Although bifermions are distributed binomially on the output modes (sublattices $q_1$ and $q_2$),  due to the splitting operation \eqref{OperatorDynamics12}, they are distributed over the Schmidt modes of both sublattices, such that  $\ket{\Psi_{N}}$ is not a superposition of two-mode coboson Fock-states $\ket{M,N-M}$. Following Appendix \ref{CobDetectionProof}, it follows that state \eqref{FiStateA} can be written as the superposition
\begin{multline}
\label{FinStateWithOrthogonals}
\hspace{-0.3cm}
\ket{\Psi_{N}} = \frac{1}{\sqrt{2^N}} \sum_{M=0}^N \sqrt{\binom{N}{M}} \left[ \sqrt{\frac{\chi_N}{\chi_M \chi_{N-M}}} \ket{M,N-M} +  \right.  \\
\left. + \sqrt{1-\frac{\chi_N}{\chi_M \chi_{N-M}}} \ket{M,N-M}^\bot \right],
\end{multline}
where $\ket{M,N-M}^\bot$ describes $M$ bifermions in the first sublattice and $N-M$ in the second, in a collective state which is orthogonal to $\ket{M,N-M}$. The populations in the two-mode Fock-state components decrease with $P$ (see Fig.\ref{CobAndPerpStatistics}, upper panel), and the contribution of their orthogonal counterpart $\ket{M,N-M}^\bot$ increases with $P$ and vanishes in the limit $P \rightarrow 0$ (see Fig.~\ref{CobAndPerpStatistics},lower panel). In this limit, cobosons exhibit a perfect bosonic behavior and the final state reads
\eq
\label{FiStateP0}
\ket{\Psi_{N}}_{P\to0} = \frac{1}{\sqrt{2^N}}\sum_{M=0}^{N} \sqrt{\binom{N}{M}} \ket{M,N-M},
\en
{\it i.e.}, cobosons are distributed binomially on the outputs as noncorrelated ideal bosons or distinguishable particles. The purity $P$ governs, therefore, to which extent the output of the beam-splitter remains in a Fock-state of cobosons.

The orthogonal Fock states $\ket{M,N-M}^\bot$ generated in the splitting process play a role analogous to the $\ket{\varepsilon_{N-1}}$ in coboson subtraction \eqref{cvecN}. Indeed,  $\ket{M,N-M}^\bot$ (coupled to the Fock-state $\ket{M,N-M}$) can be prepared by an experimental protocol \cite{KilloranCramerPlenio2014}, which projects  $\ket{\Psi_{N}}$ onto a state with a definite number of particles in each mode, leaving its internal structure unchanged, thus obtaining the projected (normalized) state
\eq
\label{ProjectiveState}
\ket{\Psi_{N}}_\text{proj} =\sqrt{\frac{\chi_N}{\chi_M \chi_{N-M}}} \ket{M,N-M} +  \nonumber\\
+ \sqrt{1-\frac{\chi_N}{\chi_M \chi_{N-M}}} \ket{M,N-M}^\bot,
\en
in full analogy with \eqref{cvecN}. Note that this post selection protocol trims the factor $\mathcal{N}^{-1} = \sqrt{\frac{1}{2^N} \binom{N}{M}}$ in Eq.~\eqref{FinStateWithOrthogonals}, due to the binomial distribution of the bi fermions on the output modes.

From Eq.~\eqref{FiStateA}, we see that no more than a single bifermion occupies the same Schmidt mode, independently of the sublattice, i.e., bifermions created by $\hat d^\dagger_{1,k_i}$ and $\hat d^\dagger_{2,k_j}$ do not share any internal Schmidt modes ($k_i \neq k_j$). Correlations due to the Pauli principle between $M$ and $N-M$ bifermions in a coboson state $\ket{N}$ are, thus,  transferred onto correlations between the $M$ and the $N-M$ bifermions which populate \emph{different} output modes after the whole process (mode-splitting and projection), as occurs with identical elementary fermions \cite{BouvrieValdesetall2016}.  With the state $\ket{\Psi_{N}}_\text{proj}$ at hand, the later mode correlation (embedded in the  component $\ket{M,N-M}^\bot$) can be tuned via the purity $P$ and used as an entanglement resource for implementing quantum information tasks \cite{BouvrieValdesetall2016,KilloranCramerPlenio2014}.

\subsection{Counting statistics}

In the 2D lattice setup in Fig.~\ref{SplittingSchedule}, the HOM-like counting statistics of an interference process of cobosons is given by the probability of finding a certain number of bifermions, say $M$, in whichever of the $S$ wells of the first sublattice, and $N-M$ in the other. It was calculated using a superposition representation in elementary bosons and fermions of the wave function \cite{TichyBouvrie2012b} (see Appendix~\ref{SuperposRep}), which allowed us to extract information on the initial two-mode Fock-state of cobosons $\ket{N_1,N_2}$ and the deviation from the ideal bosonic pattern. However, these HOM-like statistics do not carry  any detail on the resulting collective state of the bifermions or how the bifermions are distributed over the wells of the sublattices. In this section, we wish to go a step farther, and access the fine-grained details of the final state to extract information on both initial and final states by means of the occupation probability of the wells after the coboson splitting, which could be determined experimentally in this lattice setup.

The probability of finding $M$ bifermions in the first sublattice ($q_1$) and in the Schmidt modes $\{l_1,\ldots,l_M\}$ is given by (Eq.~\eqref{BifPorbProjection} in Appendix~\ref{CobDetectionProof})
\eq
\label{BifPopulation}
\mathcal{P}_\text{bif}(\{l_1,\ldots,l_M\}) 
= \frac{1}{2^N} \frac{N!}{(N-M)!} \frac{\chi_{N-M}^{[\lambda_{l_1},\ldots,\lambda_{l_M}]}}{\chi_N}\prod_{i=1}^M \lambda_{l_i},~~
\en
where $[\lambda_{l_1},\ldots,\lambda_{l_M}]$ denotes the complement of Schmidt coefficients after we have removed the coefficients $\lambda_{l_1},\ldots,\lambda_{l_M}$ from the initial distribution $\Lambda$. These probabilities constitute postselection probabilities in the sense that, for a definite number of bifermions in one sublattice $M$ (and, consequently, $N-M$ in the other), they reflect the probability that these $M$ bifermions populate the modes $\lambda_{l_1},\ldots,\lambda_{l_M}$. When the previous selection protocol, which extracts $\ket{\Psi_N}_\text{proj}$, is followed, the probability $\mathcal{P}_\text{bif}$ is modified by a factor $\mathcal{N}$, such that the selection protocol is equivalent to filtering the events with the desired population of the output modes.

Perfect bosonic behavior of cobosons involves distributions with infinitesimally small Schmidt coefficients ($\lambda_j\sim \lim_{S\to\infty}1/S \Rightarrow P\sim0$) and thus $\mathcal{P}_\text{bif} \sim 0$. Finite bifermion occupation probabilities, therefore, lead to nonideal bosonic behavior of the cobosons. The simplest case is given by the splitting of two cobosons ($N=2$) for which we obtain
\eq
\label{1BifPop2CobSplitting}
\mathcal{P}_\text{bif}(\{l\}) = \frac{\lambda_l(1-\lambda_l)}{2(1-P)},
\en
and
\eq
\label{2BifPop2CobSplitting}
\mathcal{P}_\text{bif}(\{l_1,l_2\}) = \frac{\lambda_{l_1}\lambda_{l_2}}{2(1-P)}.
\en
By experimental determination of the probability of finding one bifermion in the first sublattice in the $l$th well \eqref{1BifPop2CobSplitting}, we can determine the entanglement between the constituents if the $\lambda_l$-coefficient of the Schmidt expansion is known.  Measuring the probabilities $\mathcal{P}_\text{bif}(\{l\})$ for all $l=1,\ldots,S$, the full initial wave function is reconstructed.

Furthermore, the probabilities $\mathcal{P}_\text{bif}$ reflect the status of the final wave function, since they quantify properties of the internal structure itself. We show this by comparing \eqref{BifPopulation} with the population of the internal Schmidt mode of $\ket{M}$. To carry out such a comparison, the selection protocol is required to obtain a final state with a definite number of particles in each mode. For instance, the state that results from a two-coboson splitting plus the projection protocol with $M=1$ ($\mathcal{N}=2$), is given by
\eq
\label{2CobSplittingState}
\ket{\Psi_{2}}_\text{proj} =\sqrt{\chi_2} \ket{1,1} + \sqrt{1-\chi_2} \ket{1,1}^\bot,
\en
and the bifermion population of the Schmidt mode $\lambda_l$ of the first sublattice in this state \eqref{2CobSplittingState} is $2\mathcal{P}_\text{bif}(\{l\})$, which differs form the probability $\lambda_l$ of finding the bifermion occupying the $l$th Schmidt mode in a single-coboson state $\ket{1}$. This difference is caused by the correlation between bifermions which populate different outputs and, therefore, caused by the population of the othogonal state component $\ket{1,1}^\perp$ generated in the mode-splitting.

Indeed, the HOM-like counting statistics in the mode-splitting of cobosons, given by the sum over all possible configurations of the $M$ bifermions,
\eq
\label{BifContStat}
\mathcal{P}_\text{HOM}(M,N-M) &=&  \sum_{S\ge l_1 > \cdots > l_M \ge 1} \mathcal{P}_\text{bif}(\{l_1,\ldots,l_M\}), \nonumber \\
&=&  \frac{1}{2^N} \binom{N}{M},
\en
shows that bifermions are distributed binomially on the output modes of the beam-splitter, just like ideal bosons or distinguishable particles, without any signature of the internal structure of the cobosons. However, we show  that the impact of transitions to states $\ket{M,N-M}^\perp$ on the HOM-like counting statistics is not negligible in more complex interference scenarios.

\section{HOM Signature of states orthogonal to the Fock ladder}
\label{DoubleBS}

Mode splitting of cobosons produces the orthogonal Fock-states $\ket{N-M}^\perp$ in conjunction with the coboson Fock-states $\ket{N-M}$, Eq.~\eqref{FinStateWithOrthogonals}, and our goal in this section is to describe the distinctive signatures of the orthogonal states in HOM-like counting statistics.

\begin{widetext}

\begin{figure}[ht]
\includegraphics[scale=0.215]{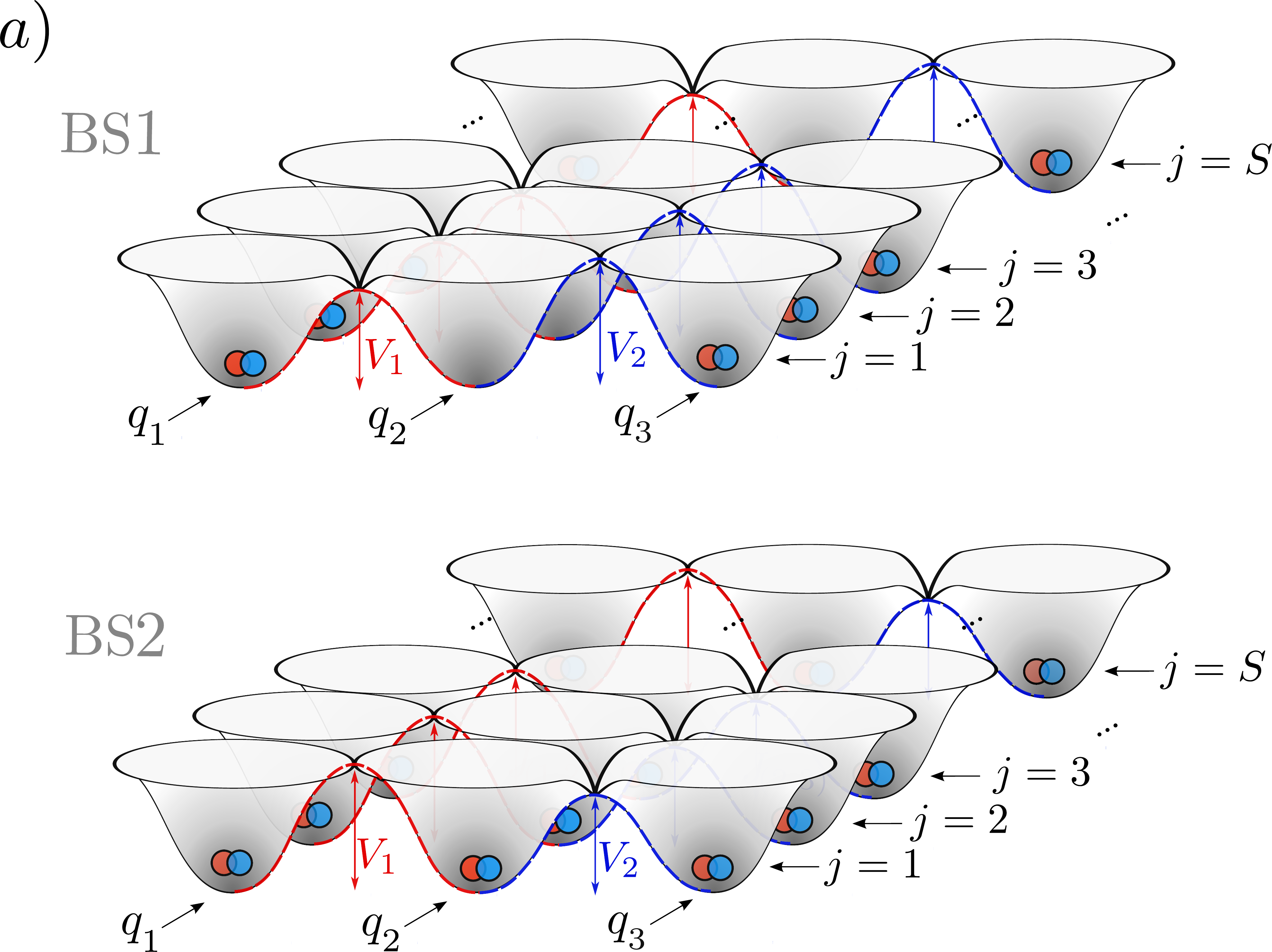} \hspace{0.6cm}
\includegraphics[scale=0.8]{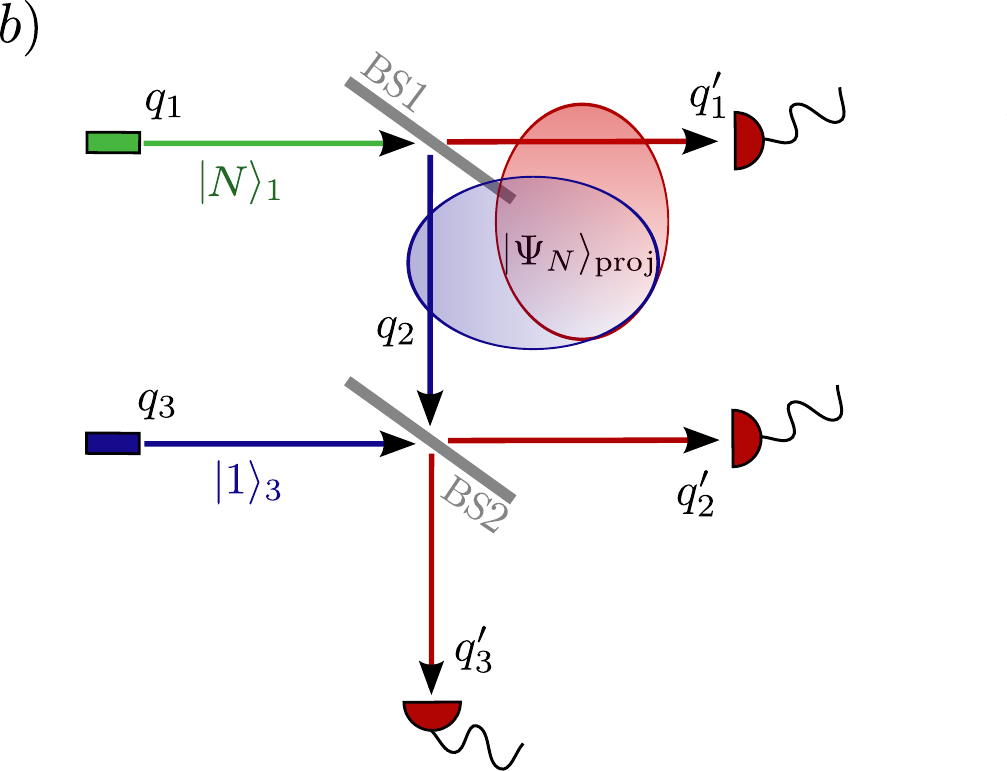}
\caption{($a$) Two-dimensional lattice setup with three external modes and controllable tunneling couplings between the $j$th wells of the sublattices. An $N$-coboson state $\ket{N}_1$, prepared in the first input $q_1$, is split by the first beam-splitter (BS$1$), such that the bifermion state in the outputs $q_1'$ and $q_2$ is given by $\ket{\Psi_{N}}$, Eq.~\eqref{FiStateA}. Then bifermions of the output $q_2$ are brought to a second beam-splitter (BS$2$) to interfere with a single coboson prepared in the third input $q_3$. In the fist dynamical process (BS$1$) the barrier between the wells of the first and the second sublattices are ramped down, $V_1\ll V_2$, where $V_1 \propto 1/J_{1,2}$ and $V_2 \propto 1/J_{2,3}$, and bifermions are allowed to cotunnel between the $j$th wells during a time interval of $t_1=\pi/(2J_{(1,2)})$. Then the barriers between the first and the second sublattices are ramped up and the barriers between the second and the third sublattices are ramped down, $V_1\gg V_2$, such that an interference process between bifermions that now populate the second sublattice and a single coboson prepared in the third takes place in BS$2$. Finally, the counting statistics of bifermions populating the outputs $q_1'$, $q_2'$, and $q_3'$ is performed after a time $t_2=\pi/(2J_{(2,3)})$. ($b$) Beam-splitter analogy of the tunnelling experiment.}	
\label{ExpRealizationFig}
\end{figure}

\end{widetext}

Our setup consists of three sublattices with $S$ wells each, see Fig.~\ref{ExpRealizationFig}$(a)$. $N$ cobosons are prepared in the first sublattice. We then couple the first and second sublattices for a time $t_1=\pi/(2J_{(1,2)})$, equivalent to a $50/50$ beam-splitter-operation. Hereafter we  couple the second and third sublattices for $t_2=\pi/(2J_{(2,3)})$ such that the bifermions in the second sublattice and the single coboson prepared in the third sublattice interfere. The dynamics of the experiment is analogous to the beam-splitter scheme depicted in Fig.~\ref{ExpRealizationFig}$(b)$. The first mode-splitting (BS1) prepares  $\ket{M,N-M}^\perp$, and we expect a deviation from the HOM-like interference of coboson Fock-states \cite{TichyBouvrie2012b} after the second mode-mixing (BS2).

In the first mode-splitting process (BS1), the initial three-mode coboson state evolves as
\eq
\label{FirstSplittingState}
\ket{N,0,1} \longrightarrow \ket{\Psi_{N}} \otimes \ket{1}_3,
\en
where $\ket{\Psi_{N}}$ is given in Eq.~\eqref{FiStateA} and
\eq
\ket{1}_3 = \sum_{i=1}^S \sqrt{\lambda_i} \hat d_{3,i}^\dagger \ket{0}.
\en
Following our discussions in the previous section, when $n_1$ bifermions are detected in the internal Schmidt modes $\lambda_{l_1},\ldots,\lambda_{l_{n_1}}$ of the first output $q_1'$, the projected state in the second mode $q_2'$, Eq.~\eqref{BifOverlapState}, contains $N-n_1$ bifermions distributed in the $S-n_1$ Schmidt modes denoted by the complement set $[\lambda_{l_1},\ldots,\lambda_{l_{n_1}}]$ in $q_2$. Since the operator $\hat c_3^\dagger$ can be written as
\eq
\label{OperatorDecomp}
\hat c_3^\dagger = \sum_{j=1}^{n_1} \sqrt{\lambda_{l_j}} \hat d_{3,l_j}^\dagger + \sum_{\substack{i=1 \\ i \neq l_1,\ldots,l_{n_1}}}^{S} \sqrt{\lambda_i} \hat d_{3,i}^\dagger ,
\en
the dynamics induced by beam-splitter BS2 can be understood as the combination of two processes: {\it (i)} a splitting process of a single bifermion, $\hat d_{3,l_j}^\dagger$, in the internal Schmidt modes $\lambda_{l_1},\ldots,\lambda_{l_{n_1}}$ of the third sublattice and ($N-n_1$) bifermions, $\hat d_{2,k_i}^\dagger$, in the (excluded) set of Schmidt modes $[\lambda_{l_1},\ldots,\lambda_{l_{n_1}}]$ in the second sublattice; and {\it (ii)} an interference process, as described in \cite{TichyBouvrie2012b}, of $N-n_1$ and $1$ bifermions in superpositions of the same $[\lambda_{l_1},\ldots,\lambda_{l_{n_1}}]$ Schmidt modes.

In Appendix~\ref{CountStatDerivation}, the HOM-like counting statistics of bifermions populating the output $q_1'$, $q_2'$, and $q_3'$, in any collective internal state is shown to evaluate to a sum of two terms representing the processes {\it (i)} and {\it (ii)},
\begin{multline}
\label{totalprobability}
\mathcal{P}_\text{HOM}(n_1,n_2,n_3) = \mathcal{Q}_\text{spl}(n_1,n_2,n_3) + \mathcal{Q}_\text{int}(n_1,n_2,n_3),
\end{multline}
where
\begin{widetext}
\eq
\label{SplitProb}
\mathcal{Q}_\text{spl}(n_1,n_2,n_3) &=& \frac{1}{2^{2N-n_1+1}} \frac{n_1}{N} \binom{N}{n_1} \left( 1- \frac{\chi_{N+1}}{\chi_N} \right) \left[ \binom{N-n_1}{n_2-1} + \binom{N-n_1}{n_3-1} \right], \\
\label{InterfProb}
\mathcal{Q}_\text{int}(n_1,n_2,n_3) &=& \frac{1}{2^{N}} \binom{N}{n_1} \left( \frac{\chi_{N+1}}{\chi_N} P_1(n_2,0) + \frac{N-n_1}{N} \left(1 - \frac{\chi_{N+1}}{\chi_N}\right) P_1(n_2,1) \right),
\en
\end{widetext}
 where $P_1(n_2,p)$ is the probability of finding $n_2$ particles in  output $q_2$ (where $p=0,1$ of them behave as fermions in the superposition representation of coboson interferences; see Appendix \ref{SuperposRep}), and $n_1+n_2+n_3=N+1$.

To compare the resulting interference pattern in the second mode-mixing (BS2), with the interference of coboson Fock-states, we perform a selection protocol (or event filtering) in the first mode-splitting process (BS1) to fix the number of particles in mode $q_2$. Thus, we are able to compare the HOM-like counting statistics of the bifermion collective interference in BS$2$ for two initial states: {\it (a)} the state $\ket{\Psi_N}_\text{proj}\ket{1}_3$ that results from the selection protocol with $M$ particles in the first mode $q_1'$, \eqref{ProjectiveState}, and {\it (b)} the usual coboson Fock-state $\ket{0,N-M,1}$. The former is given by the probabilities $\mathcal{N} \mathcal{P}_\text{HOM}(M,n_2,n_3)$  and the latter by $\mathcal{P}_\text{HOM}(n_2,n_3)$, Eq.~\eqref{UsualCobInterf} in Appendix~\ref{SuperposRep}.

\begin{figure}[ht]
\centering
\includegraphics[scale=0.7]{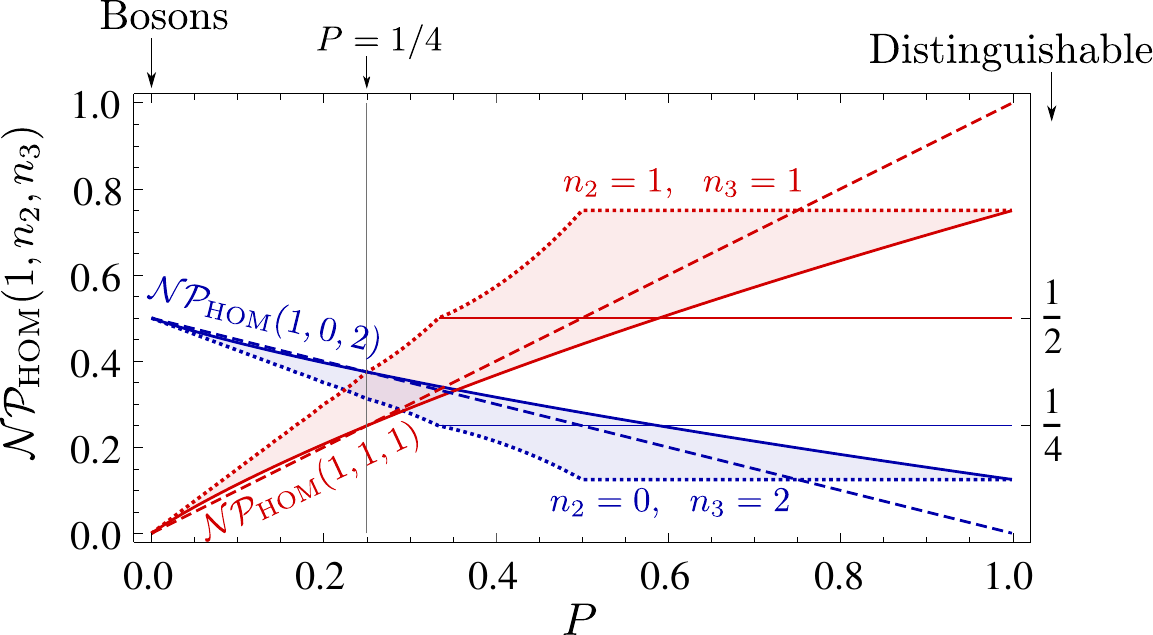}
\caption{(Color online) Postselection HOM-like interference for an initial coboson state $\ket{2,0,1}$. The probabilities $\mathcal{N} \mathcal{P}_\text{HOM}(1,n_2,n_3)$, \eqref{ProbN21} and \eqref{ProbN22}, of finding $n_2$ and $n_3$ bifermions in the modes $q_2'$ and $q_3'$, respectively, conditioned on detecting  a bifermion  in $q_1'$, are shown as a function of the purity $P$ for the peaked (solid lines) and uniform (dotted lines) distributions and are compared with the one-one coboson interference (dashed lines). Shaded areas are the possible values of the probabilities $\mathcal{N} \mathcal{P}_\text{HOM}$ for any distribution of Schmidt coefficients with purity $P$. Horizontal lines indicate the distinctive counting statistics of distinguishable particles.}	
\label{DoubleSplittForN2}
\end{figure}

For the simplest case, $N=2$ (initial state $\ket{2,0,1}$), we readily obtain the probabilities
\eq
\label{ProbN21}
\mathcal{N} \mathcal{P}_\text{HOM} (1,1,1) = \frac{3}{4} \left(1-\frac{\chi_3}{\chi_2}\right),
\en
and
\eq
\label{ProbN22}
\mathcal{N} \mathcal{P}_\text{HOM} (1,2,0) = \mathcal{N} \mathcal{P}_\text{HOM} (1,0,2) = \frac{1}{4} \left(1+3\frac{\chi_3}{\chi_2}\right).
\en
In Fig.~\ref{DoubleSplittForN2}, we show the upper and lower bounds of these probabilities (given by the peak $\Lambda^\text{peak}$ and uniform $\Lambda^\text{uni}$ distributions of the Schmidt coefficients \cite{TichyBouvrie2012a}) as a function of the purity $P$. The populations for the post-selected interference may, in general, differ from the ``one-one'' coboson interference given by $\mathcal{P}_\text{HOM} (1,1) = 1-\chi_2=P$ and $\mathcal{P}_\text{HOM} (2,0) = \chi_2/2=(1-P)/2$ (dashed lines in Fig.~\ref{DoubleSplittForN2}), and they are strictly different for $0<P<1/4$ and $3/4<P<1$. These differences are caused by the correlations generated in the splitting process between bifermions which populate $q_1'$ and $q_2$, and which are described by  $\ket{1,1}^\perp$. The interfering bifermions in BS2 carry some memory of the pre-selected bifermion on mode $q_1'$, and, therefore, although there are two interfering bifermions in BS$2$, the maximum degree of the normalization factor in Eqs.~\eqref{ProbN21} and \eqref{ProbN22} is $3$. For $P<1/4$ (vertical line), the postselection interference enhances the deviation from the bosonic pattern.

For an initial state $\ket{N,0,1}$ with arbitrary $N \ge 1$, the probability of obtaining the last bifermion in modes $q'_2$ and $q'_3$, when all $N$ bifermions incident in $q_1$ are selected in the first output $q'_1$, is obtained as $\mathcal{N} \mathcal{P}_\text{HOM} (N,0,1) = \mathcal{N} \mathcal{P}_\text{HOM} (N,1,0) = 1/2$,
since BS2 merely splits the single bifermion population in $q_3$. When all $N$ incident particles with probability $1/2^N$ impinge on the second beam-splitter after splitting in BS$1$, the usual coboson interference \cite{TichyBouvrie2012b} takes place in BS$2$, with
\begin{multline}
\label{UsualInterference}
\mathcal{N}\mathcal{P}_\text{HOM} (0,n_2,N-n_2+1) = \frac{\chi_{N+1}}{\chi_{N}} P_1(n_2,0) + \\
+ \left(1-\frac{\chi_{N+1}}{\chi_{N}}\right) P_1(n_2,1),
\end{multline}
(c.f. Eq.~\eqref{UsualCobInterf} with $M=0$).

However, when $M$ particles (with $0<M<N$) are detected in the first output $q_1'$, the splitting process in BS1 generates transitions to states given by $\ket{\Psi_N}_\text{proj}$, Eq.~\eqref{ProjectiveState}, and deviations from the usual interference patterns of coboson Fock-states in BS$2$ are statistical signatures of the mode-correlated state $\ket{M,N-M}^\perp$. A postselection interference process allow us, therefore, to characterize the bosonic behavior of the whole systems: While the HOM-like counting statistics of the cobosons in the initial Fock-state $\ket{0,N-M,1}$, \eqref{UsualCobInterf}, depends on the normalization ratio $\chi_{N-M+1}/\chi_{N-M}$, the postselection interference probabilities $\mathcal{N} \mathcal{P}_\text{HOM}(M,n_2,n_3)$ are functions of $\chi_{N+1}/\chi_{N}$.

\begin{figure}[ht]
\centering
\includegraphics[scale=0.7]{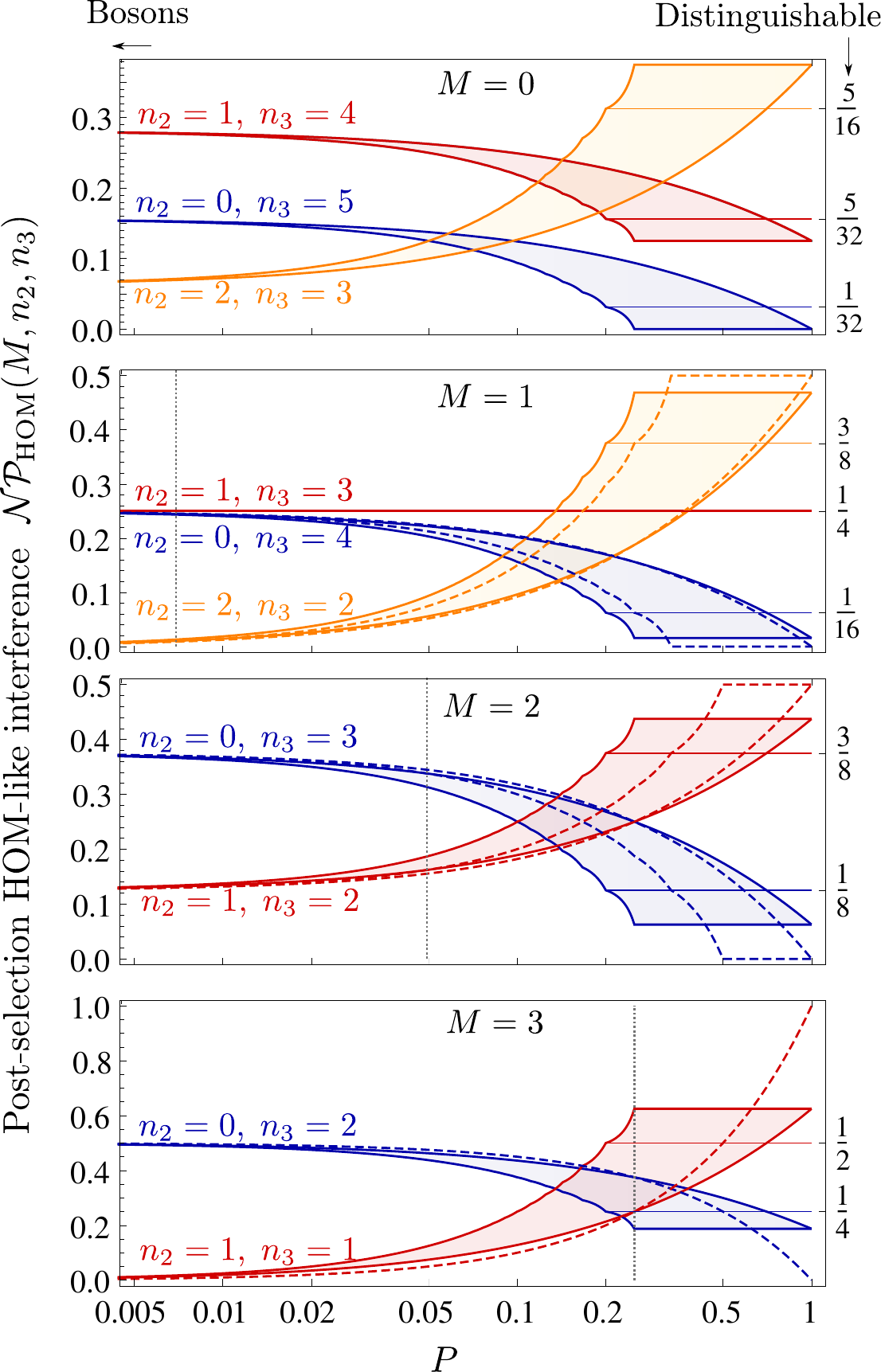}
\caption{(Color online) Postselection HOM-like interference: Probabilities $\mathcal{N}\mathcal{P}_\text{HOM}(M,n_2,n_3)$ of an initial coboson state $\ket{4,0,1}$ ($N=4$) and for different numbers $M$ of (postselected) bifermions in the first output $q_1'$. For $P=0$, perfect bosonic behavior emerges irrespective of the distribution of Schmidt coefficients. The uniform and peaked distributions of Schmidt coefficients (solid lines) limit the possible values (shaded areas) of the probabilities $\mathcal{N}\mathcal{P}_\text{HOM}$, for finite $P$. The usual collective interference $\mathcal{P}_\text{HOM}(n_2,n_3)$ for initial coboson Fock-states $\ket{0,4-M,1}$ is plotted by dashed lines, and the statistics of distinguishable particles is represented by horizontal lines. The upper part of the figure shows the limiting $(1,4)$-cobosons interfere case of Eq.~\eqref{UsualInterference}. In the range $P<P_\text{enh}$ (vertical lines) the deviation from the bosonic pattern is strictly larger for a post-selected HOM-like interference than for the usual coboson interference.}	
\label{CountingStatisticsN4}
\end{figure}

The deviation from the ideal bosonic behavior is negligible when constituents of composites are maximally entangled ($P\approx0$); cobosons behave as perfect bosons and the HOM-like counting statistics in BS$2$ will be the same as for elementary bosons $\mathcal{N}\mathcal{P}_\text{HOM}^\text{bosons}(M,n_2,n_3) = P_1(n_2,0)$. When the normalization ratio fulfills $\chi_{N+1}/\chi_N = 1/(N+1)$, the probability, \eqref{totalprobability}, matches the distinguishable-particle case,
\eq
\mathcal{N}\mathcal{P}_\text{HOM}^\text{dist}(M,n_2,n_3)= \frac{1}{2^{N-M+1}} \binom{N-M+1}{n_2}.
\en
Thus, by varying the purity $P$, a transition between fully bosonic and fully distinguishable behavior may be implemented experimentally, even though the bifermions always remain indistinguishable. 

Postselection HOM-like counting statistics, $\mathcal{N}\mathcal{P}_\text{HOM}(M,n_2,n_3)$, for $N=4$ are shown in Fig.~\ref{CountingStatisticsN4} (solid lines) as a function of the purity for the peaked $\Lambda^\text{peak}$ and the uniform $\Lambda^\text{uni}$ Schmidt distributions. As a general trend, the deviations from the ideal bosonic statistics increase with $P$. The larger the number of particles detected  in the first output mode ($M$), the more strongly the interference pattern differs from the usual interference pattern of coboson Fock-states (dashed lines). For small values of the purity,
\eq
\label{BoundForEnhance}
\sqrt{P}< \sqrt{P_\text{enh}} = \frac{M}{(N-1)(N-M+1)},
\en
the deviation from ideal bosonic behavior is always larger than in the usual coboson interference (see left-hand-side of the vertical dotted lines in Figs.~\ref{DoubleSplittForN2} and \ref{CountingStatisticsN4}). Thus, for cobosons with purity in the range of \eqref{BoundForEnhance}, one ensures that by adding successive beam-splitters the effects of compositeness are enhanced further and further.

\begin{figure}[ht]
\centering
\includegraphics[scale=0.8]{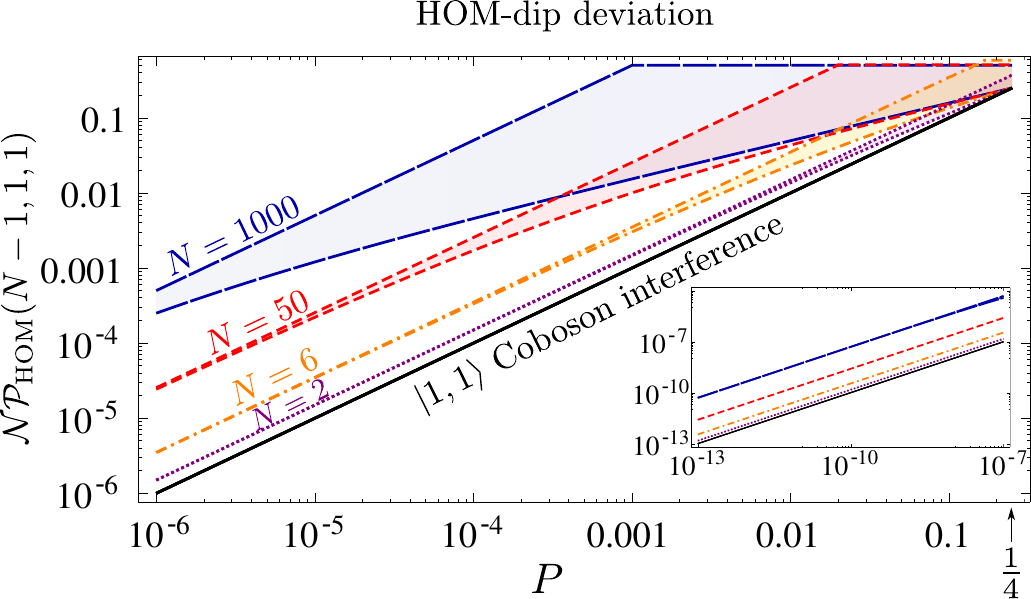}
\caption{(Color online) Bounds on the probability $\mathcal{N}\mathcal{P}_\text{HOM}(N-1,1,1)$ (HOM-dip) of finding a coincident event in the second beam-splitter (BS$2$) after an ($N-1$) bifermion postselection in the first output $q_1'$ of BS$1$, as a function of $P$, for $N=2, 6, 50, 1000$. For $N=1$, the deviation of the HOM-dip corresponds to the usual ``one-one'' coboson interference (solid line). Inset: The range of very small purities.}	
\label{Enhancement}
\end{figure}

By means of the post-selected interference process described above, the deviation of the coboson interference pattern from the ideal bosonic pattern can be increased by several orders of magnitude with respect to simpler procedures and scenarios \cite{TichyBouvrie2012b}, which facilitates the detection of signatures of compositeness. This is shown, e.g., by the finite probability of finding a coincident event in an $M=N-1$ post-selected coboson interference, where just two bifermions interfere in BS$2$. In Fig.~\ref{Enhancement}, we plot the deviation  from the HOM-dip of an $N-1$ post-selected coboson interference, given by the probability $\mathcal{N}\mathcal{P}_\text{HOM}(N-1,1,1)$ (which vanishes exactly for the interference of two elementary bosons). The deviation from the ideal increases with the number of pre-selected ($N-1$) bifermions, such that the probability of finding coincident events in the output $q_2'$ and $q_3'$ increases by more than one order of magnitude for $N=50$ and almost three for $N=1000$ (dashed lines in Fig.~\ref{Enhancement}) with respect to the usual ``one-one'' coboson interference (solid line).

\section{Conclusion}
\label{Conclusion}

Splitting dynamics of cobosons manifest the transition to states not described by the usual Fock-states of cobosons. The particles are distributed binomially on the external splitting modes, just like elementary bosons or distinguishable particles, but the collective structure of the constituents on the internal Schmidt modes changes dramatically in the splitting process, and nonideal bosonic operators $\hat c^\dagger$ lead to states with particles in the output modes that are anticorrelated in their internal states.

The counting statistics for composite bosons made by superpositions of ultracold atoms in lattice wells remains difficult to obtain experimentally, but the bifermion counting statistics in the mode-splitting process, which depends 
on the internal structure of the resulting composite-particle-state, reveals that this splitting process of composite bosons generates correlations between particles that populate different output modes. Signatures of these mode-correlations are also found in the HOM-like counting statistics after the beam-splitter-like dynamics induced by two consecutive splitting arrangements, with the advantage that no information of the internal structure of the composite system is required to extract properties of the wave function such as the purity and the normalization ratio.

While naturally occurring composite bosons, like atoms and molecules, are directly detectable, the electron-state purities of trapped ultracold atoms  are typically prohibitively low, of the order of $10^{-12}$ \cite{ChudzickiOkeEtal2010,RombutsVanNeck2002}. The deviations from the bosonic pattern will be roughly of the same order, such that atomic HOM experiments \cite{LopesImanalievEtal2015} and interference with BECs of atoms \cite{JeltesEtal2007} are not sensitive to the compositeness of the atoms. The methods we present here, i.e.~scenarios concatenating a splitting process and postselection interference, enhance the signatures of compositeness by orders of magnitude and relax the precision required in experiments to detect deviations from the ideal bosonic behavior.

In mixtures of ultracold Fermi gases, the interaction between fermionic atoms (and consequently, the entanglement) can be tuned by means of external fields to create molecular bound states (Feshbach resonances), which leads to the Bose-Einstein condensation of molecules \cite{GreinerRegalJin2003}. The coboson theory successfully applies to these many diatomic composites  with non-trivial composite signatures \cite{BouvrieTichyRoditi2016}; the coboson state $\ket{N}$ describes essentially the Bose-Einstein condensate at zero temperature, and the particle statistics, such as the condensate fraction of fermion pairs, depend on the entanglement (and, thus, on the purity $P$). The impressive progress in the control of few-fermion systems in ultracold atom experiments \cite{SerwaneZurnEtal2011,ZurnSerwane2012,ZurnJochim2013} has allowed us to implement experimentally the Hubbard model with interacting fermionic atoms in double-well systems \cite{Jochim2015}. The HOM-like counting statistics of ultra-cold atoms is feasible \cite{AndrewsTownsendEtal1997,LopesImanalievEtal2015} and interference of molecular BECs has been observed \cite{KohstallRiedlEtal2011}. Hence, the preparation of states \eqref{InState}, and the observation of full beam-splitter-like dynamical processes in a lattice seem to be challenging, yet rewarding goals for ultracold atom experiments.

\begin{appendix}

\section{Coboson interference and superposition representation}
\label{SuperposRep}

The collective interference of cobosons does not allow us to use an operator evolution of the form of \eqref{OperatorDynamics12}, since the essential bifermion on-site interactions are neglected in this description. However, the superposition representation in elementary bosons and fermions permits to derive how the bifermions are distributed over the output modes of a beam-splitter in an interference process of coboson Fock-states \cite{TichyBouvrie2012b}. Despite this representation's being a powerful tool for the physical interpretation of interference processes, it does not incorporate any sign of the collective structure of the coboson constituents over the $S$ internal Schmidt modes. It is a high-level description that does not take into account the resulting state but the consequences in the HOM counting statistics. We summarize in the following the main ingredients of this approach, which are used for the calculation of $\mathcal{P}_\text{HOM}(n_1,n_2,n_3)$ in Appendix~\ref{CountStatDerivation}.

States with at least one coboson in each external mode can be represented as
\eq
\label{TwoModesCobState}
\ket{N_1,N_2} \simeq \sum_{p=0}^{\text{Min}[N_1,N_2]} \sqrt{\omega_p} \ket{\phi_p(N_1,N_2)},
\en
where the states $\ket{\phi_p(N_1,N_2)}$, which represent $2p$ bifermions that behave as fermions and $N_b= N_1+N_2-2p$ as bosons, are orthogonal. Their corresponding weights are given by
\eq
\label{weight}
w_p =
{N_1 \choose p }  {N_2 \choose p } ~ \frac{p!}{\chi_{N_1} \chi_{N_2} } ~\Omega(\{ \underbrace{2,\dots, 2}_p,\underbrace{1,\dots ,1}_{N_b}\}).
\en 
The probability of finding $m$ bifermions in the first lattice after an ($N_1$-$N_2$)-coboson interference process is the sum of the resulting probabilities from the different contributions of the state $\ket{N_1,N_2}$ in the superposition representation,
\eq
\mathcal{P}_\text{HOM}(m,N_\text{tot}-m) = \sum_{p=0}^{\text{Min}\{m,N_\text{tot}-m\}} \omega_p P(m,p),
\en
where $N_\text{tot}= N_1+N_2$.  $P(m,p)$ is the probability of finding $m$ bifermions in the first lattice after the interference process, where $p$ of them (with $p\le m$) behave as fermions and $m-p$ as bosons. This is given by the amplitude
\eq
P(m,p) = 
|\mathcal{A}\left(N_1-p,N_2-p,m - p,N_\text{tot}-m-p\right)|^2,
\en
which can be evaluated for a perfect beam-splitter dynamics ($T=R=1/2$) with the methods presented in Refs.~\cite{TichyLimEtal2011,TichyTierschEtal2012,LaloeMullin2012} as
\begin{align}
\mathcal{A}(n_1,n_2,m_1,m_2) = i^{m_1-n_1} \frac{2^{\frac{1}{2} (2+m_1+m_2)}}{m_1-n_1!} \sqrt{\frac{m_1!n_2!}{m_2!n_1!}} \times \nonumber \\
\times {}_2F_1[1 + m_1, 1 + m_1 + m_2 - n_1; 1 + m_1 - n_1; -1],
\end{align}
where ${}_2F_1$ is the ordinary (or Gauss) hypergeometric function.

The particular case that we use is given by the interference between a single coboson and $N-n_1$ cobosons (initial state $\ket{0,1,N-n_1}$ in the three-sublattice model)
\eq
\label{UsualCobInterf}
\mathcal{P}_\text{HOM} (n_2,n_3) &=& \frac{\chi_{N-n_1+1}}{\chi_{N-n_1}} P_1(n_2,0)  \\
&+& \left(1-\frac{\chi_{N-n_1+1}}{\chi_{N-n_1}}\right) P_1(n_2,1),~~ \nonumber
\en
where $n_3=N-n_1+1-n_2$ and
\eq
P_1(n_2,p) = ~~~~~~~~~~~~~~~~~~~~~~~~~~~~~~~~~~~~~~~~~~~~~~~~~~~~~~ \\
|\mathcal{A}\left( N-n_1-p , 1-p , n_2-p , N-n_1+1-n_2-p \right)|^2 \nonumber .
\en

\section{Proof of the splitting state decomposition}
\label{CobDetectionProof}

In this appendix we prove \eqref{FinStateWithOrthogonals} in the text. For this purpose, we first compute the projection
\eq
\ket{d_{l_1},\ldots,d_{l_M}}_{1~1}\braket{d_{l_1},\ldots,d_{l_M}}{\Psi_{N}}_\text{f} = \nonumber \\
= \ket{d_{l_1},\ldots,d_{l_M}}_{1} \ket{\psi_{N-M}}_{2}
\en
where
\eq
\label{BifState}
\ket{d_{l_1},\ldots,d_{l_M}}_{1} = \prod_{i=1}^M \hat d_{1,l_i}^\dagger \ket{0},
\en
and $\ket{\Psi_{N}}$ is the final state given by Eq.~\eqref{FiStateA}. $\ket{\psi_{N-M}}_{2}$ is the resulting state in the second output mode $q_2$ after the detection of $M$ bifermions in the $\{l_1,\ldots,l_M\}$ Schmidt modes of the first output $q_1'$. The only nonvanishing term of superposition \eqref{FiStateA} describes $M$ particles in the first mode, such that the state in $q_2$ is given by
\begin{widetext}
\eq
\ket{\psi_{N-M}}_{2} &=& \frac{1}{\sqrt{2^N N! \chi_{N}}} \sum_{\substack{k_1,\ldots,k_{N}=1 \\ k_1 \neq \cdots \neq k_{N}}}^S ~  {}_1\bra{d_{l_1},\ldots,d_{l_M}} \prod_{j=1}^N \sqrt{\lambda_{k_j}} \left( \hat d_{1,k_j}^\dagger + \hat d_{2,k_j}^\dagger \right) \ket{0,0}\\
&=& \sqrt{\frac{1}{2^N N! \chi_{N}}} \sum_{\substack{k_1,\ldots,k_{N}=1 \\ k_1 \neq \cdots \neq k_{N}}}^S \left(\prod_{j=1}^N \sqrt{\lambda_{k_j}}\right) {}_1\bra{d_{l_1},\ldots,d_{l_M}} \binom{N}{M} \prod_{r=N-M+1}^N \hat d_{1,k_r}^\dagger \prod_{s=1}^{N-M} \hat d_{2,k_s}^\dagger \ket{0,0}. \nonumber
\en
Since there are $M!$ possible combination to match the $M$-index $k_r$ and $l_i$, the above state reads
\eq
\label{BifOverlapState}
\ket{\psi_{N-M}}_{2}  = \frac{1}{\sqrt{2^N N! \chi_N}} \frac{N!}{(N-M)!} \prod_{i=1}^M \sqrt{\lambda_{l_i}} \sum_{\substack{k_1,\ldots,k_{N-M}=1 \\ k_1 \neq \cdots \neq k_{N-M} \neq l_1 \neq  \cdots \neq l_M}}^S \prod_{j=1}^{N-M} \sqrt{\lambda_{k_j}} \hat d_{2,k_j}^\dagger \ket{0}_2.
\en
The probability of finding $M$ bifermions in the first sublattice ($q_1$) and in the Schmidt modes $\{l_1,\ldots,l_M\}$, Eq.\eqref{BifPopulation} of the main text, is given by the probability amplitude of the projection on the state $\ket{d_{l_1},\ldots,d_{l_M}}_{1}$,
\eq
\label{BifPorbProjection}
\mathcal{P}_\text{bif}(\{l_1,\ldots,l_M\}) = \left|{}_1\braket{d_{l_1},\ldots,d_{l_M}}{\Psi_{N}}_\text{f}\right|^2
= {}_2\braket{\psi_{N-M}}{\psi_{N-M}}_{2}.
\en

An $M$-coboson state in the first mode is written in terms of state \eqref{BifState} as
\eq
\ket{M}_1 = \frac{1}{\sqrt{M! \chi_M}}\sum_{\substack{l_1,\ldots,l_{M}=1 \\ l_1 \neq  \cdots \neq l_M}}^S \left( \prod_{i=1}^M \sqrt{\lambda_{l_i}}\right) \ket{d_{l_1},\ldots,d_{l_M}}_1.
\en
Using Eq.~\eqref{BifOverlapState}, the state that results from the projection of such an $M$-coboson state in the first mode onto state $\ket{\Psi_N}_\text{f}$ is given by $\ket{M}_{1~1}\braket{M}{\Psi_N}_\text{f}$, where
\begin{multline}
\label{CobOverlapState}
{}_1\braket{M}{\Psi_N}_\text{f}  = \frac{1}{\sqrt{M!\chi_M}}\sum_{\substack{l_1,\ldots,l_{M}=1 \\ l_1 \neq  \cdots \neq l_M}}^S {}_{1}\bra{d_{l_1},\ldots,d_{l_M}} \prod_{i=1}^M \sqrt{\lambda_{l_i}}  \ket{\Psi_{N}} = \\
=\frac{1}{\sqrt{2^N N! \chi_N M!\chi_M}} \frac{N!}{(N-M)!} \sum_{\substack{l_1,\ldots,l_{M}=1 \\ l_1 \neq  \cdots \neq l_M}}^S \left( \prod_{i=1}^M \sqrt{\lambda_{l_i}} \right)^2 \sum_{\substack{k_1,\ldots,k_{N-M}=1 \\ k_1 \neq \cdots \neq k_{N-M} \neq l_1 \neq  \cdots \neq l_M}}^S \prod_{j=1}^{N-M} \sqrt{\lambda_{k_j}} \hat d_{2,k_j}^\dagger \ket{0}_2.
\end{multline}
Since
\eq
\sum_{\substack{k_1,\ldots,k_{N-M}=1 \\ k_1 \neq \cdots \neq k_{N-M} \neq l_1 \neq  \cdots \neq l_M}}^S \prod_{j=1}^{N-M} \sqrt{\lambda_{k_j}} \hat d_{2,k_j}^\dagger \ket{0}_2 = \frac{(N-M)!}{N!}  \left(\prod_{i=1}^M \frac{\hat d_{2,l_i}}{\sqrt{\lambda_{l_i}}} \right) \sum_{\substack{k_1,\ldots,k_{N}=1 \\ k_1 \neq \cdots \neq k_{N}}}^S \prod_{j=1}^{N} \sqrt{\lambda_{k_j}} \hat d_{2,k_j}^\dagger \ket{0}_2,
\en
Eq.~\eqref{CobOverlapState} reads
\begin{multline}
{}_1\braket{M}{\Psi_N}_\text{f}  = \frac{1}{\sqrt{2^N N! \chi_N M!\chi_M}}  \sum_{\substack{l_1,\ldots,l_{M}=1 \\ l_1 \neq  \cdots \neq l_M}}^S \left( \prod_{i=1}^M \sqrt{\lambda_{l_i}} \hat d_{2,l_i} \right) \sum_{\substack{k_1,\ldots,k_{N}=1 \\ k_1 \neq \cdots \neq k_{N}}}^S \prod_{j=1}^{N-M} \sqrt{\lambda_{k_j}} \hat d_{2,k_j}^\dagger \ket{0}_2 = \\
= \frac{1}{\sqrt{2^N M!\chi_M}}  \sum_{\substack{l_1,\ldots,l_{M}=1 \\ l_1 \neq  \cdots \neq l_M}}^S \left( \prod_{i=1}^M \sqrt{\lambda_{l_i}} \hat d_{2,l_i} \right) \frac{1}{\sqrt{N! \chi_N}} \sum_{\substack{k_1,\ldots,k_{N}=1 \\ k_1 \neq \cdots \neq k_{N}}}^S \prod_{j=1}^{N-M} \sqrt{\lambda_{k_j}} \hat d_{2,k_j}^\dagger \ket{0}_2,
\end{multline}

\end{widetext}
that is,
\eq
\label{singlemodeoverlap}
{}_1\braket{M}{\Psi_N}_\text{f} = \frac{1}{\sqrt{2^N}}\frac{\left( \hat c_2 \right)^M }{\sqrt{M!\chi_M}} \ket{N}_2.
\en
Therefore, the state which results from apply the $M$-coboson projection onto the initial state, \eqref{FiStateA}, is
\eq
\label{OneOutputProjStateApp}
\ket{M}_{1~1}\braket{M}{\Psi_{N}}_\text{f} = \frac{1}{\sqrt{2^N}} \frac{\left( \hat c_2 \right)^{M}}{\sqrt{M!\chi_{M}}} \ket{M,N}.
\en

With the above Eq.~\eqref{OneOutputProjStateApp} at hand, and using
\eq
\hat c^\dagger \ket{N} = \sqrt{N+1} \sqrt{\frac{\chi_{N+1}}{\chi_N}} \ket{N+1},
\en
the projection of an ($N-M$)-coboson state in the second output onto Eq.~\eqref{OneOutputProjStateApp} reads
\begin{multline}
\ket{M,N-M}\braket{M,N-M}{\Psi_N}_\text{f} = \\
=\frac{1}{\sqrt{2^N M!\chi_M}} \bra{N-M} \hat c^M \ket{N} \ket{M,N-M} \\
= \sqrt{\frac{1}{2^N} \binom{N}{M} \frac{\chi_N}{\chi_M \chi_{N-M}}} \ket{M,N-M}.
\end{multline}
Therefore, the state $\ket{\Psi_N}_\text{f}$ can be written as the superposition given in Eq.~\eqref{FinStateWithOrthogonals}.

\noindent $\blacksquare$

\section{Derivation of the HOM-like counting statistics of the double beam-splitter setup}
\label{CountStatDerivation}

We derive the probabilities $\mathcal{P}_\text{HOM}(n_1,n_2,n_3)$ (Eqs. \eqref{totalprobability}, \eqref{SplitProb}
and \eqref{InterfProb}) of finding $n_1$, $n_2$ and $n_3$ bifermions in the outputs $q_1'$, $q_2'$, and $q_3'$, respectively, in the setup in Fig.~\ref{ExpRealizationFig}. By a probe, the atoms pairs which occupy the wells (or Schmidt modes) in each of the three sublattices,  $q_1'$, $q_2'$ and $q_3'$, are detected in the experiment. Since we are not interested in particular the wells in which the fermion pairs are located -- just the number of occupied wells in each sublattice $n_1$, $n_2$, and $n_3$ -- we simply sum all possible bifermions configurations along the Schmidt modes of the outputs. Thus, the order in which the particles are detected in the measurement process is irrelevant.

When $M$ bifermions are detected in the Schmidt modes $\{l_1,\ldots,l_M\}$ of the first sublattice $q_1'$ after the splitting, the resulting state in the sublattice $q_2$ is given by Eq.~\eqref{BifOverlapState}. Therefore, the state in the sublattices $q_2$ and $q_3$ before the second interference process in BS$2$, once the detection of the $M$ bifermions in the first sublattice is performed, can be written using \eqref{OperatorDecomp} as
\eq
\ket{d_{l_1},\ldots,d_{l_{n_1}}}_{1~1}\braket{d_{l_1},\ldots,d_{l_{n_1}}}{\Phi}_\text{BS1} = ~~~~~~~~~~~~~~ \nonumber \\
= \ket{d_{l_1},\ldots,d_{l_{n_1}}}_{1~1}\braket{d_{l_1},\ldots,d_{l_{n_1}}}{\Psi_{N}}_\text{f} \otimes \ket{1}_3  \nonumber \\
= \ket{d_{l_1},\ldots,d_{l_{n_1}}}_{1} \ket{\psi_{N-n_1}}_{2}\otimes \ket{1}_3  \nonumber \\
= \ket{d_{l_1},\ldots,d_{l_{n_1}}}_{1}  (\ket{\phi_A}_{2,3} + \ket{\phi_B}_{2,3})
\en
where
\begin{widetext}
\eq
\label{InStateBefoerInterf}
\ket{\phi_A}_{2,3} =
\frac{1}{\sqrt{2^N N! \chi_N}} \frac{N!}{(N-{n_1})!} \prod_{i=1}^{n_1} \sqrt{\lambda_{l_i}} \sum_{\substack{k_1,\ldots,k_{N-{n_1}}=1 \\ k_1 \neq \cdots \neq k_{N-{n_1}} \neq l_1 \neq  \cdots \neq l_{n_1}}}^S \prod_{j=1}^{N-{n_1}} \sqrt{\lambda_{k_j}} \hat d_{2,k_j}^\dagger \left( \sum_{i=1}^{n_1} \sqrt{\lambda_{l_i}} \hat d_{3,l_i}^\dagger \right) \ket{0,0}_{2,3}, \\
\ket{\phi_B}_{2,3} =
\frac{1}{\sqrt{2^N N! \chi_N}} \frac{N!}{(N-{n_1})!} \prod_{i=1}^{n_1} \sqrt{\lambda_{l_i}} \sum_{\substack{k_1,\ldots,k_{N-{n_1}}=1 \\ k_1 \neq \cdots \neq k_{N-{n_1}} \neq l_1 \neq  \cdots \neq l_{n_1}}}^S \prod_{j=1}^{N-{n_1}} \sqrt{\lambda_{k_j}} \hat d_{2,k_j}^\dagger \left(   \sum_{\substack{i=1 \\ i \neq l_j}}^{S} \sqrt{\lambda_i} \hat d_{3,i}^\dagger \right) \ket{0,0}_{2,3}.
\en
\end{widetext}
States $\ket{\phi_A}_{2,3}$ and $\ket{\phi_B}_{2,3}$ are orthogonal, and hence, after the second beam-splitter dynamics, the probability of finding $n_2$ and $N-n_2-n_1+1$ bifermions in sublattices $q_2'$ and $q_3'$, respectively, is the sum of the probabilities of both contributions, $\mathcal{Q}_A$ and $\mathcal{Q}_B$.

State $\ket{\phi_A}_{2,3}$ constitutes a superposition of single bifermion states such that one bifermion ($\hat d_{3,l_j}^\dagger$) is distributed along the Schmidt modes $\lambda_{l_1},\ldots,\lambda_{l_{n_1}}$ of the third output mode and $N-n_1$ bifermions ($\hat d_{2,k_i}^\dagger$) are distributed along the set of Schmidt modes $[\lambda_{l_1},\ldots,\lambda_{l_{n_1}}]$  of the second external mode. The set $[\lambda_{l_1},\ldots,\lambda_{l_{n_1}}]$ is the distribution $\Lambda$ to which we have removed the coefficients $\lambda_{l_1},\ldots,\lambda_{l_{n_1}}$. Since bifermions of both external modes, $\hat d_{3,l_j}^\dagger$  and $\hat d_{2,k_i}^\dagger$, do not share any internal Schmidt modes, the beam-splitter dynamics for an initial state $\ket{\phi_A}_{2,3}$ is that given by a splitting process. The probability of such a splitting process ($n_1$ bifermions have been detected on the Schmidt modes $\{l_1,\ldots,l_{n_1}\}$ of the first sublattice) is given by
\begin{widetext}
\eq
\mathcal{Q}_A =  \frac{1}{2^{2N-n_1}} \frac{N!}{(N-{n_1})!} \frac{1}{2} \left[\binom{N-n_1}{n_2-1}+ \binom{N-n_1}{N-n_1-n_2} \right] \sum_{j=1}^{n_1} \prod_{i=1}^{n_1} \lambda_{l_j} \lambda_{l_i} \frac{\chi_{N-n_1}^{\tilde \Lambda}}{\chi_N},
\en
where we have taken into account that the bifermions are distributed binomially in the output modes $q_2'$ and $q_3'$, as in Eq.~\eqref{BifContStat}, and that the state
\eq
\label{NormalizedStateLessLambdas}
\frac{1}{\sqrt{(N-n_1)! \chi_{N-n_1}^{\tilde \Lambda}}} \sum_{\substack{k_1,\ldots,k_{N-{n_1}}=1 \\ k_1 \neq \cdots \neq k_{N-{n_1}} \neq l_1 \neq  \cdots \neq l_{n_1}}}^S \prod_{j=1}^{N-{n_1}} \sqrt{\lambda_{k_j}} \hat d_{k_j}^\dagger \ket{0}
\en
is normalized to unity. Since
\eq
\sum_{1 \leq l_1<l_2<\cdots<l_{n_1}\leq S} ~ \sum_{j=1}^{n_1} \prod_{i=1}^{n_1} \lambda_{l_j} \lambda_{l_i} \chi_{N-n_1}^{\tilde \Lambda} =  \frac{n_1}{n_1! N} \left(\chi_N-\chi_{N+1}\right),
\en
summing over all bifermion configurations leads to \eqref{SplitProb}
\eq
\mathcal{Q}_\text{spl}(n_1,n_2,n_3) = \sum_{1 \leq l_1<l_2<\cdots<l_{n_1}\leq S} \mathcal{Q}_A ~ = ~ \frac{1}{2^{2N-n_1+1}} \frac{n_1}{N} \binom{N}{n_1} \left( 1- \frac{\chi_{N+1}}{\chi_N} \right) \left[ \binom{N-n_1}{n_2-1} + \binom{N-n_1}{N-n_2-n_1} \right],
\en

State $\ket{\phi_B}_{2,3}$ describes a single bifermion, created by $\hat d_{3,l_j}^\dagger$, and $N-n_1$ bifermions, created by $\hat d_{2,k_i}^\dagger$ in the third and the second external modes, respectively. All of them are in a superposition of the internal Schmidt modes given by the set $\tilde \Lambda = [\lambda_{l_1},\ldots,\lambda_{l_{n_1}}]$. Thus, the beam-splitter dynamics for an initial state $\ket{\phi_B}_{2,3}$ is the one given by a coboson interference process \cite{TichyBouvrie2012b} as shown in Appendix~\ref{SuperposRep}. Using Eqs.~\eqref{NormalizedStateLessLambdas}, \eqref{TwoModesCobState}, and \eqref{weight}, we have that
\eq
\mathcal{Q}_B =  \frac{1}{2^{N}} \frac{N!}{(N-{n_1})!} \frac{\prod_{i=1}^{n_1} \lambda_{l_j}}{\chi_N}  \left[ \Omega^{\tilde \Lambda}(\{ \underbrace{1,\dots ,1}_{N-n_1+1}\}) P_1(n_2,0) + (N-n_1) \Omega^{\tilde \Lambda}(\{2, \underbrace{1,\dots ,1}_{N-n_1-1}\}) P_1(n_2,1) \right].
\en
Thus, summing over all bifermion configurations,
\eq
\sum_{1 \leq l_1<l_2<\cdots<l_{n_1}\leq S} ~ \left(\prod_{i=1}^{n_1} \lambda_{l_i}\right) \Omega^{\tilde \Lambda}(\{ \underbrace{1,\dots ,1}_{N-n_1+1}\}) =  \frac{\chi_{N+1}}{n_1!} ,
\en
\eq
\sum_{1 \leq l_1<l_2<\cdots<l_{n_1}\leq S} ~ \left(\prod_{i=1}^{n_1} \lambda_{l_i}\right) \Omega^{\tilde \Lambda}(\{2, \underbrace{1,\dots ,1}_{N-n_1-1}\}) =  \frac{\chi_N-\chi_{N+1}}{n_1!N} ,
\en
we find \eqref{InterfProb}
\eq
\mathcal{Q}_\text{int}(n_1,n_2,n_3) = \sum_{1 \leq l_1<l_2<\cdots<l_{n_1}\leq S} \mathcal{Q}_B ~ = ~ \frac{1}{2^{N}} \binom{N}{n_1} \left( \frac{\chi_{N+1}}{\chi_N} P_1(n_2,0) + \frac{N-n_1}{N} \left(1 - \frac{\chi_{N+1}}{\chi_N}\right) P_1(n_2,1) \right).
\en

\end{widetext}

\end{appendix}

\section*{Acknowledgments}

We thank Fernando de Melo for carefully reading the manuscript. P.A.B. belongs to Andalusian group FQM-020 and gratefully acknowledges support from the Conselho Nacional de Desenvolvimento Cient\'ifico e Tecnol\'ogico do Brasil through a BJT {\it Ci\^encia sem Fronteiras} Fellowship, and from Spanish project grant no. FIS2014-59311-P (cofinanced by FEDER). KM acknowledges support from the Villum Foundation


\end{document}